\input amstex

\documentstyle{amsppt} 

\magnification1200 \TagsOnRight 
\NoBlackBoxes
\pagewidth{30pc} \pageheight{47pc} 

\def\today{10 October 1997}

\topmatter
\title A note on area variables in Regge calculus\endtitle
\author John W. Barrett, Martin Ro\v cek, Ruth M. Williams \endauthor

\date\today\enddate

\address
Department of Mathematics,
University of Nottingham,
University Park,
Nottingham,
NG7 2RD, UK
\endaddress
\email jwb\@maths.nott.ac.uk \endemail
\address
 Institute for Theoretical Physics
State University of New York at Stony Brook
Stony Brook NY 11794-3840 USA
\endaddress
\email rocek\@insti.physics.sunysb.edu\endemail
\address
DAMTP,
Silver Street,
Cambridge 
CB3 9EW, UK
\endaddress
\email rmw7\@damtp.cam.ac.uk \endemail
 
\abstract
We consider the possibility of setting up a new version of Regge calculus in
four dimensions with areas of triangles as the basic variables rather than the
edge-lengths. The difficulties and restrictions of this approach are discussed.
\endabstract                 
 
\thanks Preprint no.~ITP-SB-97-59 \endthanks
\endtopmatter

\document 
     In three dimensions, the topological invariant for three-manifolds
constructed by Turaev and Viro [1] reduces in the semi-classical limit to the
expression written down more than twenty years earlier by Ponzano and Regge [2].
In the case of a very fine triangulation of the manifold, Ponzano and Regge had
pointed out an intriguing connection between their state sum and the Feynman
path integral for three-dimensional quantum gravity using the Regge calculus
action [3]. A pre-requisite for this was the very obvious one that in the
Turaev-Viro and Ponzano-Regge expressions, `colourings' or angular momentum
variables are assigned to the edges (links) of the triangulation, and in Regge
calculus the edge lengths are indeed the variables playing the role of the
metric tensor in the continuum theory.

      An extension of the Turaev-Viro type of state sum to four dimensions
[4,5] involved assigning the colourings or variables to the
two-dimensional faces of the triangulation. The relationship between this
work and the loop representation of quantum gravity has been explored by
Rovelli [6], who suggested that it should be related to a modified version
of Regge calculus in four dimensions, in which the independent variables
would be the areas of the triangles rather than the edge lengths. Prompted
by this remark, we set out to explore this possibility.

More recently, the idea of the areas of triangles as variables in state
sum models of quantum gravity featured in [7,8], and the relation of these
models to continuum Lagrangians was explored in [9,10].

Rovelli's ideas fitted in with earlier work by M\"akel\"a [11] who
considered a version of Regge calculus using the areas of triangles as
coordinates, based on a modification of the Ashtekar formalism. M\"akel\"a 
noted that in Regge calculus, there are constraints among the area
variables, and concentrated on taking account of these constraints so that
his formalism still corresponded to using edge lengths [12]. In contrast,
in the present work our intention is to explore the possibility of
treating the areas as the fundamental and independent variables.

     The fact that a four-simplex has the same number of triangular faces as
edges is very suggestive, and as Rovelli claims, one would expect to be able to
invert the relationship between $A_t$, the area of triangle t, and the squares of the edge-lengths, $s_i$, in order to express the Regge action
 $$ I(s_i) = \sum_t   A_t(s_i) \epsilon_t(s_i)     \tag 1$$
in terms of the areas $A_t$.
 Here $\epsilon_t$ is the deficit angle at triangle t. Then one could look for stationary points of the action, leading to discrete
solutions of Einstein's equations. However, as we shall see, there are a number
of difficulties with this approach.
 
     Firstly let us look at the problem of inverting the relationship between
lengths and areas for a given four-simplex. Label the vertices $1,2,\ldots,5$; 
let
$s_{ij}$ be the length-squared of the edge joining vertices $i$ and $j$ and 
$A_{ijk}$ the
area of the triangle with vertices $i,j,k$. Then
 $$A^2_{ijk} = \left( {2(s_{ij} s_{jk} + s_{ik} s_{jk} + s_{ij} s_{ik}) - s_{ij}^2 - 
s_{jk}^2 - s_{ik}^2}\right) /16.    \tag 2$$

As Philip Tuckey pointed out, there  are pairs of four-simplexes,
with squared edge-lengths $(1,1,1,1,1,1,1,1,1,a)$  which have the same triangle
areas, but with different values of $a$ either side of $2$ [13]. This follows from
the fact that the $A^2_{ijk}$ are quadratic in $a$ and are all a maximum
at $a=2$. This corresponds to a geometry in which
some of the triangles are right-angled.  Thus there would have to
be a restriction on the class of edge-lengths considered, corresponding to a
restriction on the class of metrics, to avoid this two-fold ambiguity.
This can be done by restricting to some neighbourhood of the regular 
four-simplex.

      There is another problem in going from edge-length variables to areas.
Although there is a match between the numbers of these for a single 
four-simplex, this does not  hold for a collection of four-simplexes
joined together. For example, consider two four-simplexes meeting on a common
tetrahedral face. This complex has 14 edges but 16 triangles. 
For any closed four-manifold, the number of triangles is greater than or equal
to $4/3$ times the number of edges. Thus even if the edge-lengths are restricted
so that the transformation to the set of areas is injective, there will be
some complicated constraint equations among the areas which arise in this
way.

     To explore this problem further, consider  two four-simplexes as
separate entities, each with the areas of its ten faces specified. Then,
assuming a restriction to geometries where the transformation between areas and
edge lengths is 1-1, we solve for the edge lengths of the two four-simplexes.
Now return to the situation where the two four-simplexes are joined on a common
tetrahedral face and require that the areas of the common triangles be equal.
Since a tetrahedron has six edges and four triangles, the
areas certainly do not determine the edge lengths of the tetrahedron uniquely, 
so we can envisage
a bizarre situation where the edge lengths of the common tetrahedron have
different values, depending on which four-simplex they are considered to be
part of. 
 
For example, suppose that for one four-simplex, all the squared areas
are 1, and the squared edge lengths are all $4/\sqrt3$. Another four-simplex
shares a tetrahedron, so four of its squared areas must be 1. Suppose that of
the remaining six, two more are 1, but the four others are 625/768. The
solution for the squared edge lengths in this four-simplex is that two opposite
sides of the common tetrahedron have squared-length 3, while the remaining
eight edges have squared-length 25/12. (The Jacobian of the transformation is
non-zero for both four-simplexes). We see the impossibility of interpreting
the edge lengths as real physical quantities in the usual sense.  However since the edge lengths are well-defined within each simplex, it is still possible to define the metric in terms of them in the usual way within each simplex.
 
Finally we need to see what types of discrete solutions of Einstein's
equations we obtain, for the restricted class of metrics avoiding the 2-fold ambiguity. Then, the deficit
angles are determined unambiguously by the areas, since each 4-simplex has well-determined hyperdihedral angles. The action is then
 
$$    I(A_s) =  \sum _{t}   A_t \epsilon_t(A_s)                  \tag3$$
where the sum is over all triangles $t$.
Variation with respect to $A_u$ gives
$$    0 = \epsilon_u(A_s) + \sum_t A_t {\partial \epsilon_t \over\partial A_u}
       = \epsilon_u(A_s) + \sum_t A_t  {\partial  \over\partial A_u}( 2 \pi - \sum_{j} \theta_t^j)                \tag4$$
where $\theta_t^j$ is the internal hyperdihedral angle at triangle t in
four-simplex j, and the second sum is over those $j$ containing $t$.

Using the chain rule and interchanging the order of summation
over triangles and over four-simplices, we obtain
 $$    0 = \epsilon_u(A_s) - \sum_{j}
          \sum_{t} A_t \sum_i
 {\partial  \theta_t^j \over\partial  s_i} { \partial  s_i \over \partial  A_u}              \tag5$$
where $t$ now runs over the triangles in 4-simplex $j$, and $i$ runs over the edges for $j$, with $s_i$ the length squared of $i$. 

Now Regge proved [3] that for the triangles in a four-simplex,
 
$$    \sum_{t} A_t {\partial  \theta_t^j    \over  \partial s_i} = 0  ,      \tag6$$
and then, since the Jacobian $\partial  s_i / \partial  A_u$ is non-singular, the variational equation reduces to $\epsilon_u = 0$ for all $u$. In Regge calculus, this condition would mean the space is locally flat.
However in our theory the deficit angle $\epsilon$ does not measure the
holonomy of a connection and so there is no such simple conclusion to
be drawn.

\head Conclusions \endhead
 Making the areas of triangles independent variables for Regge calculus is 
possible, but the price for this is that there appears to be no well-defined metric geometry. Work is continuing on this puzzling aspect of the formalism. The action can still be defined, as the dihedral angles are
determined in each simplex, and these are all that are required to define
the deficit angles. A calculation shows that the variational equations imply 
that these deficit angles are all zero, but this does not imply that the 
solutions are flat metrics.

\head {Acknowledgements}\endhead
 
We are grateful to Francis Archer and Philip Tuckey for helpful discussions. Martin Ro\v cek was supported by NSF grant no.~PHY9722101.

 \enddocument